\documentclass{article}
\usepackage{epsfig,latexsym,graphics}
\def\circa#1{\,\raise.3ex\hbox{$#1$\kern-.75em\lower1ex\hbox{$\sim$}}\,}
\newcommand{\ov}{{\cal O}}

\newcommand \bra {\langle}
\newcommand \ket {\rangle} \newcommand{\be}{\begin{equation}}
\newcommand{\ee}{\end{equation}} \newcommand{\ben}{\begin{displaymath}}
\newcommand{\een}{\end{displaymath}} \newcommand{\ba}{\begin{eqnarray}}
\newcommand{\ea}{\end{eqnarray}} \newcommand{\ban}{\begin{eqnarray*}}
\newcommand{\ean}{\end{eqnarray*}} \newcommand{\cro}{\dagger}

\newcommand{\pt}{\mbox{$p$}_{\perp }}

\def\eeeq{\end{eqnarray}}

\def\beeq{\begin{eqnarray}}
\def\eeeq{\end{eqnarray}}

\def\bom#1{{\mbox{\boldmath $#1$}}}
\def\to{\rightarrow}

\def\nn{\nonumber}

\def\ID{1 \kern -.45 em 1}
\def\bom#1{{\mbox{\boldmath $#1$}}}

\begin{document}

\vspace{1.cm}

{\centering

{\Large\bf The Importance of Weak Boson Emission at LHC}

\vspace{1.cm}

{\bf \large Paolo Ciafaloni}

{\it INFN - Sezione di Lecce, \\Via per Arnesano, I-73100 Lecce, Italy \\
E-mail: paolo.ciafaloni@le.infn.it}
\vspace{0.4cm}

{\bf \large Denis Comelli}

{\it INFN - Sezione di Ferrara, \\Via Saragat 1, I-43100 Ferrara, Italy\\
E-mail: comelli@fe.infn.it}

}

\vspace{0.3cm}

\begin{abstract}
We point out that gauge bosons emissions should be
carefully estimated when considering LHC observables, since real $Ws$ and $Zs$
contributions can dramatically change cross sections with respect to tree
level values.
 Here we consider observables 
that are fully inclusive
respect to soft gauge boson emission
and where a certain number of nonabelian
isospin charges in initial and/or final states are detected.
We set up a general formalism to evaluate leading, all order resummed
electroweak corrections and we consider the phenomenologically relevant
case of third family quark production at the LHC. In the case of $b\bar{t}$
production we find that, due to the interplay between strong and weak
interactions, the production cross section 
can become an order of magnitude bigger than the tree level value.
\end{abstract}

\section{Introduction}
It is by now well established that one loop electroweak corrections are not
sufficient to keep under control Standard Model predictions at the TeV
scale. The reason for this is the sharp growth with energy of these kind of
corrections, that reach the 10 \% level at 1 TeV. More in detail, this
growth is related to the infrared structure of the theory, so that one loop
contributions are proportional to $\log^2\frac{s}{M_{W,Z}^2}$, the gauge bosons
masses $M_{W,Z}$ acting as infrared regulators \cite{IR}. 
In order to cope with the
expected precision of hadronic (LHC)  and leptonic (ILC)
 colliders, fixed-order
one and  two loop corrections  \cite{fixedSuda}  and resummation of leading
effects  \cite{resumSuda} have been
considered by various groups in the last few years. 
Broadly speaking, two kinds of observables have been considered. In first
place the exclusive observables where only virtual electroweak effects need to be
considered \cite{fixedSuda,resumSuda}. 
In second place, observables including $W,Z,\gamma$ emissions
have been considered \cite{EWBN}. A prototype of these kind of observables is $e^+
e^-\to hadrons$: the two final jets are detected, while any other object in
the final state is summed over, including the final decay products of $Ws$
and $Zs$ \cite{EWBN}. In this case, the collider provides two initial nonabelian
charges, and due to this fact the outcome is surprising: even though fully
inclusive, the cross section is sensitive to the infrared cutoff $M_W$ and
affected by big  $\log^2\frac{s}{M_{W}^2}$ terms. This effect, baptized
``Bloch-Nordsieck violation'', has been shown to occur only in broken gauge
theories, including the abelian case \cite{abelian}. Recently,
electroweak evolution equations, which are the analogous of QCD DGLAP
equations, have been derived \cite{col}.

The aim of this work is to considered another class of observables, which
we might call ``partially inclusive''. That is, gauge boson radiation in
is still summed over, however we retain the possibility to observe
nonabelian charges in both  the initial and/or  final states. 
The cross sections we consider are, generally speaking, identified by 4
hard partons, 2 in the initial and 2 in the final state. However in order
to compute the leading logs related to virtual and real $W,Z,\gamma$
emission we can reduce to 2 or 3 relevant legs. Suppose for instance that
the flavor of leg 4 is undetected (see fig.1) This could be the case if we
consider leg 4 to be a final jet and we do not isolate the jet's
flavor. Then, by unitarity, since we are summing over real and virtual
corrections \cite{EWBN}, we need only to consider the electroweak
corrections related to legs 1,2,3. This is showed in fig. 1 where the
overlap matrix with the remaining three legs is depicted.

\begin{figure}
   \centering
   \includegraphics[width=16cm]{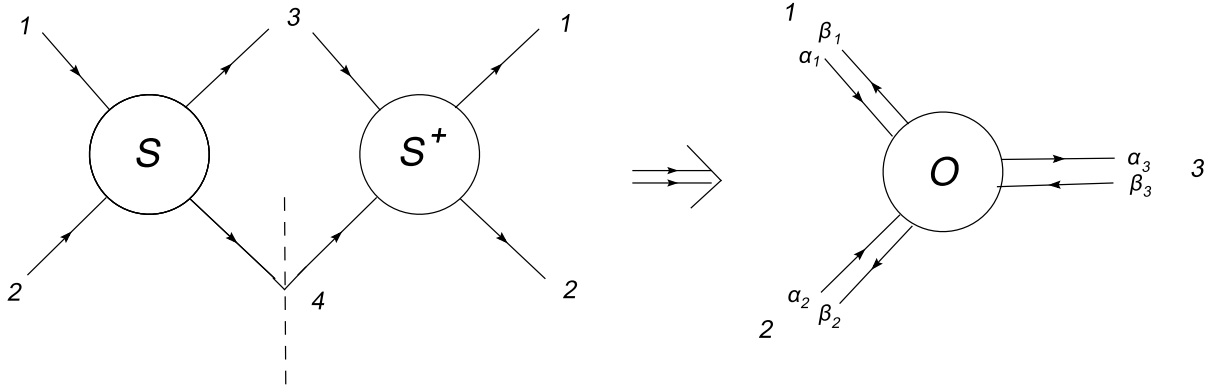}
   \caption{Diagrammatic Isospin description of the overlap matrix
   with three isospin charged legs and one inclusive final state.}
  \end{figure}

\section{General formalism}
Our starting point, as in previous works, is the SU(2) isospin structure of
 the overlap matrix, defined in
terms of the S-matrix as follows \cite{EWBN} ($\alpha_i,\beta_i$ 
are the isospin indices):
\be
\bra \beta_1\beta_2\dots \beta_n|S^\cro S|\alpha_1\alpha_2\dots \alpha_n\ket
=\ov_{\beta_1\alpha_1,\beta_2\alpha_2,\dots\beta_n\alpha_n}
\ee
and the observable cross sections are related to the above
definition by:
\be
d\sigma_{\alpha_1\alpha_2\dots\alpha_n}=\ov_{\alpha_1\alpha_1,
\alpha_2\alpha_2,\dots\alpha_n\alpha_n}
\ee
Notice that we use a ``generalized S matrix'' formalism, such that all the
 states over which we are inclusive appear on the left of S and all the
 detected nonabelian charges $1,2...n$ appear on the right. So for instance
 a detected final (outgoing) antiquark is seen as an initial (ingoing)
 quark state. 
Namely, for the case $n=2$  this  means that we describe  
cross sections with two electroweak charges  in the initial
states or systems with   one charged particle in the initial and one in 
the final state,
 or the case of  two final charges.

The SU(2) generators $t^a, a=1,2,3\quad i=1,2,\dots n$ act on the overlap
matrix as follows
\be
(t^a_i \ov)_{\beta_1\alpha_1,\dots\beta_n\alpha_n}=
\sum_{\delta_i}t^a_{\alpha_i\delta_i}
 \ov_{\beta_1\alpha_1,\dots\beta_i\delta_i,\dots\beta_n\alpha_n}
\qquad
(t'^a_i \ov)_{\beta_1\alpha_1,\dots\beta_n\alpha_n}=
\sum_{\gamma_i}t^a_{\beta_i\gamma_i}
 \ov_{\beta_1\alpha_1,\dots\gamma_i\alpha_i,\dots\beta_n\alpha_n}
\ee
where the generators $t^a$ depend on the representation of the considered
$i-th$ particle.

It is convenient to define the isospin generator referred
to a single leg $i$ as $T_i\equiv t_i+t'_i$  \cite{EWBN}.
Since we consider energy scales of the order of 1 TeV and beyond, we take all particles to
be massless.
In other words  we work in the high energy limit in which the $SU(2)\otimes U(1)$
symmetry is recovered; then the  overlap matrix  is invariant under a symmetry
transformation:
\be\label{sym} {T}_{tot}^a\equiv\sum_i { T}_i^a\qquad
\exp[\alpha^a { T}^a_{tot}]\ov=
\exp[\bom{\alpha}\cdot \bom{ T}_{tot}]\ov=\ov\quad
\Rightarrow \quad \bom{T}_{tot}\ov=0
\ee
This property allows to write the overlap matrix as a sum of projectors with
definite isospin properties and gives various relations between the apriori
independent cross sections (see next sections). 

The dressing of the hard overlap matrix $\ov^H$ to obtain the evolved one
$\ov$ through exchange of virtual and real soft $W$ quanta 
 is described by the external line insertion of the eikonal current:
\begin{equation}
\label{eikcur}
{\bom J}^{\mu }(k)=g_w\;\sum_{i=1}^{n} {\bom T}_i\,
\frac{p_{i}^{\mu}}{p_i\cdot k} \;,
\end{equation}
$k$ being the momentum of the emitted soft gauge boson, $p_i$ the i-th
leg momentum and $g_w$ the SU(2) gauge coupling. 
Notice that the part of  the current
proportional to $g'$ is absent altogether because of the cancellation 
of the abelian components for inclusive observables \cite{EWBN}.

By squaring the eikonal current one obtains, 
in the limit where all invariants are of the same order $2p_i\cdot p_j\approx s$, 
the
insertion operator:
\be
I(k)=g_w^2\frac{p_1p_2}{(kp_1)(kp_2)}\sum_{i< j}^n\bom{T}_i\cdot\bom{T}_j
\ee

The resummed expression for the overlap matrix is finally given by the
following expression involving the insertion operator:
\be\label{res}
\ov(s)=\exp[L_W \sum_{i< j}^n\bom{T}_i\cdot\bom{T}_j]\ov^H
\ee
where we have defined the eikonal radiation factor for $W$ exchange:
\be
L_W(s)=\frac{g_w^2}{2}\int_M^E\frac{d^3\bom{k}}{2\omega_k(2\pi)^3}
\frac{2p_1p_2}{(kp_1)(kp_2)}=\frac{\alpha_w}{4\pi}\log^2\frac{s}{M^2},
\quad\alpha_w=\frac{g_w^2}{4\pi}
\ee

It is useful to rewrite (\ref{res}) by using 
\be
\sum_{i<j}^n\bom{T}_i\cdot\bom{T}_j=\frac{1}{2}\sum_{i=1}^n
\bom{T}_i\cdot(\bom{T}_{tot}-\bom{T}_i)
=-\frac{1}{2}\sum_i \bom{T}_i^2
\ee
where we used $\bom{T}_{tot}\ov=0$, so that
\be\label{dress}
\ov(s)=\exp\left[-\frac{1}{2}L_W \sum_{i=1}^n\bom{T}_i^2\right]\ov^H
\ee
Eqn. (\ref{dress}) shows the single particle property of inclusive
emission, i.e. the fact that corrections are calculated by considering an
exponential factor for each leg in a definite total isospin state. In the
following we systematically adopt the procedure of writing the overlap
matrix as a sum of projectors on total isospin eigenstates and then applying
(\ref{dress}) to obtain the ``EW BN'' corrected overlap matrix. 
The hard overlap matrix is therefore first decomposed as follows:
\be\label{expansion}
\ov^H=\sum_{t_1,t_2\dots t_n} O^H_{t_1t_2\dots t_n}{\cal{P}}_{t_1t_2\dots t_n}
\ee
where $\ov,{\cal{P}}_{t_1t_2\dots t_n}$ are operators acting on the $n$
external legs indices, and $O_{t_1t_2\dots t_n}$ are the coefficients of
the expansion. The projectors satisfy, by definition:
\be\label{prop}
\bom{T}_j{\cal{P}}_{t_1t_2\dots t_n}=t_j{\cal{P}}_{t_1t_2\dots
  t_n},j=1\dots n
\qquad
\bom{T}_{tot}{\cal{P}}_{t_1t_2\dots t_n}=0
\ee
Then we apply (\ref{dress}) in order to obtain the all order resummed
values. Due to property (\ref{prop}) this is particularly simple, since it
amounts to the substitution:
\be\label{subs}
 O^H_{t_1t_2\dots t_n}\to O_{t_1t_2\dots
   t_n}(s)=\exp\left[-\frac{1}{2}L_W(s) 
\sum_{i=1}^n t_i(t_i+1)\right]O^H_{t_1t_2\dots t_n}
\ee
In next section we give the explicit form of the projection operators for
the cases of two and three external legs.

To end with, we want to compare the above describe ``BN EW'' corrections
with ``Sudakov EW'' corrections, 
 i.e. EW corrections given only by the virtual
contributions without weak bosons emissions. The latter depend on how the
observable is defined, namely on which cutoff is decided  on real photon
emission:  a certain degree of inclusiveness on photons is
mandatory in order to render the observable infrared finite. For
definiteness and in order to compare with the BN corrections, we choose for
the photon a cutoff of the order of the weak scale; the result is an
effective SU(2) $\otimes$ U(1) theory with all gauge bosons at a common
mass $M_W\approx M_Z$ \cite{resumSuda}.
In this limit Sudakov corrections are in fact rather simple: the resummed cross section
is obtained from the hard one by multiplying each external leg by an
exponential factor:
\be\label{suda}
\sigma^{Sud}(s)=\exp [-L_W(s) \sum_i(t_i(t_i+1)+y_i^2\tan^2\theta_W )]\sigma_H
\ee
 where $\theta_W$ is
the Weinberg angle, $t_i$ is the i-th leg isospin and $y_i$ its hypercharge
\cite{resumSuda}. 
Despite the similarities between (\ref{subs}) and (\ref{suda}), the
inclusive (BN) and exclusive (Sudakov) case are of course very different and give rise to
significantly different patterns of radiative corrections. Namely:
\begin{itemize}
\item
in (\ref{suda}) $t_i$ is the external leg isospin (e.g., $\frac{1}{2}$ for
a fermion) while in  (\ref{subs}) $t_i$ is obtained by
composing two single-leg isospins (see fig. 1)
\item
no correction proportional to $y^2$ is present in the ``BN'' inclusive
case, since contributions proportional to the U(1) coupling $g_y$ cancel
out \cite{EWBN}.
\item
There is a factor 2 of difference in the argument of the exponential
(compare (\ref{suda}), (\ref{subs})).
\item
while Sudakov corrections always depress the tree level cross section, 
BN ones can be negative or positive (see
section 4).
\end{itemize}

\section{The case of two and three external legs}
In this section we give the explicit forms of the projectors in the case of
two and three external legs. This allows to calculate the EW BN
corrections by simply inserting the appropriate values of the hard cross
sections, as we explain in next section. Of course not all of the possible
values of $t_1,t_2\dots t_n$ appearing in (\ref{expansion}) are allowed. In
fact the total isospin must be $0$  due to isospin invariance, so
for instance in the case of two legs there is no ${\cal{P}}_{10}$ term,
since no isospin invariant can be constructed from an isospin 0 and an
isospin 1 objects. In the following we present tables with the allowed
values for $t_1\dots t_n$. The coefficients $O_{t_1t_2\dots t_n}$ are also
called ``form factors'' since they are $s$-dependent and
receive the exponential factor (\ref{subs}). Here we limit ourselves to
fermions, antifermions and transverse gauge bosons in the external
legs. Therefore in the following by ``boson'' we always mean ``transversely
polarized (weak) gauge boson''.

We now consider the ``two external legs'' case. Notice that by this we do
not mean that only two external particles are present. Rather, we mean a
process with an arbitrary number of external particles, but in which 
only two non abelian weak charges are detected. Therefore, the process $gg\to
q\bar{q}$ belongs to this category since gluons do not carry weak charges;
also the process $e^+e^-\to jets+X$ falls in this case since this process
is fully inclusive in the final state: no weak charge is singled out. The
case of two initial external charged legs has been widely discussed
from various point of view:
in \cite{EWBN} the BN violation was put in  evidence and more refined 
studies were done in \cite{col} including the next to leading 
corrections.
Here we extend the same procedure to  processes with
{\it two external charged legs} irrespective of their position of
initial or final states.  

The possible values for the isospin of the two external legs labeled by
$t_1,t_2$ are given by (f.f.=form factors):

\vspace{.3cm}

\noindent\begin{tabular}{|c|c|c| |c|c|c| |c|c|c|}
\hline
fermion & fermion & Number of f.f. & fermion & boson & 
Number of f.f. & boson & boson &Number of f.f. \\
\hline
$t_1=0$ & $t_2=0$&  & $t_1=0$ &$ t_2=0$ &  & $t_1=0$ &$ t_2=0$ &  \\
$t_1=1$ & $t_2=1$& 2 & $t_1=1$ & $t_2=1$ & 2 & $t_1=1$ &$ t_2=1$& 3\\
$$ & $$& & $$ & $$ & & $t_1=2$ &$ t_2=2$&\\
\hline
\end{tabular}

\vspace{.3cm}

The explicit form of the projection operators is given below for the
relevant case of fermions and (transverse) gauge bosons:

\begin{itemize}
\item two external fermions
\ba
\ov(\alpha_1,\beta_1;\alpha_2,\beta_2)
=O_{00}\;\delta_{\alpha_1\beta_1}\delta_{\alpha_2\beta_2}
+O_{11}\;t^a_{\alpha_1\beta_1}t^a_{\alpha_2\beta_2}
\label{2f}\ea
\item one fermion and one external gauge boson
\ba
\ov(\alpha_1,\beta_1;a_2,b_2)
=O_{00}\;\delta_{\alpha_1\beta_1}\delta_{a_2b_2}
+O_{11}\;t^a_{\alpha_1\beta_1}T^a_{a_2b_2}
\ea
\item two external  gauge  bosons
\ba
\ov(a_1,b_1;a_2,b_2)
=O_{00}\;\delta_{a_1b_1}\delta_{a_2b_2}
+O_{11}\;T^a_{a_1b_1}T^a_{a_2b_2}+
O_{22}\;{\cal{P}}_2(a_1b_1;a_2b_2)
\ea
\end{itemize}
where $t^a(T^a)$ are the SU(2) generators in the fundamental (adjoint)
 representation and
 the isospin 2 projector is defined by:
\be
{\cal{P}}_2(a_1,b_1;a_2,b_2)=\frac{1}{4}\left[\{T^c,T^d\}_{b_1a_1}
\{T^c,T^d\}_{b_2a_2}-\frac{16}{3}\delta _{b_1a_1}\delta_{b_2a_2}\right]
\ee
The case with three external particles charged under $SU(2)$  is more
complicated because the product of three isospins generates many
invariant with definite total isopin.
As shown in the table below 
in order to describe a system with three external fermion in the
fundamental representation five gauge invariant form
factors are needed; a system with two fundamental fermions and one boson
(in the adjoint representation ) needs six form factors and for a system
with one fermion plus two bosons we have to write nine form factors.

\vspace{0.3cm}

\begin{tabular}{|c|c|c|c||c|c|c|c|}
\hline
fermion & fermion & fermion &Number of f.f.
 &fermion & fermion & boson & Number of f.f.
\\ \hline
$t_1=0$ &$ t_2=0$ &$ t_3=0$& &$t_1=0$ & $t_2=0$ &$ t_3=0$&\\
$t_1=1$ &$ t_2=0$ &$ t_3=1$&$5$ &$t_1=1$ &$ t_2=0$ &$ t_3=1$&$6$\\
$t_1=0$ & $t_2=1$ &$ t_3=1$& &$t_1=0$ &$ t_2=1$ &$ t_3=1$&\\
$t_1=1$ &$ t_2=1$ & $t_3=0,1$&&$t_1=$1 &$ t_2=1$ &$ t_3=0,1,$2 &\\
\hline
\end{tabular}

\vspace{0.1cm}

\begin{tabular}{|c|c|c|c|}
\hline
 fermion & boson & boson & Number of f.f.
\\\hline
$t_1=0$ &$ t_2=0$ &$ t_3=0$&\\
$t_1=$1 &$ t_2=0$ & $t_3=1$&\\
$t_1=0$ & $t_2=1$ &$ t_3=1$& $9$\\
$t_1=$1 &$ t_2=1$ &$ t_3=0,1,$2 &\\
$t_1=0$ &$ t_2=2$ &$ t_3=2$&\\
$t_1=1$ &$ t_2=2$ &$ t_3=1,$2 &\\
\hline
\end{tabular}

\vspace{0.2cm}

For the case with three bosons there are 15 form factors.
The form factors are gauge invariant combinations of
  physical cross sections (that correspond to the diagonal elements of
  the overlap matrix).
Reversing the problem, any physical cross section is a combination of
  form factors.
 It is interesting  to note that the form factors
  $\ov_{ijk...}$ whose sum $i+j+k+...=$ {\it odd number }  do not
  contribute to physical cross sections.
In practice this  means that the degrees of freedom of the overlap
  matrix projected on the physical space of the cross sections are
  diminished.
In the above examples, with three external particles the form factors
  $\ov_{111}$ and $\ov_{122}$ are unphysical. The  final result  is that, at the level of
  physical cross sections, the system with
  three fermions has 4 degrees of freedom: once we know  4 cross sections any
  other one is fixed  as a  combination of these ones.
The system of two fermions and one boson has 5 degrees of freedom and
  finally the one fermion and two bosons only 7.

The isospin decomposition of the overlap  matrix in
the various  cases is given below.
 Fermionic and antifermionic legs are treated on equal grounds, so
for instance the case of one antifermionic and two fermionic legs belongs
to the ``3 fermionic legs'' case. 
\begin{itemize}
\item 3 fermionic legs
\ba\label{3f}
\ov(\alpha_1,\beta_1;\alpha_2,\beta_2;\alpha_3,\beta_3)
&=&O_{000}\; \delta_{\alpha_1\beta_1}\delta_{\alpha_2\beta_2}\delta_{\alpha_3\beta_3}
+O_{101}\; t^a_{\alpha_1\beta_1}\delta_{\alpha_2\beta_2}t^a_{\alpha_3\beta_3}+
O_{111}\; f_{abc}t^a_{\alpha_1\beta_1}t^b_{\alpha_2\beta_2}t^c_{\alpha_3\beta_3}
\nn\\
&+&
O_{011}\; \delta_{\alpha_1\beta_1}t^a_{\alpha_2\beta_2}t^a_{\alpha_3\beta_3}
+O_{110}\; t^a_{\alpha_1\beta_1}t^a_{\alpha_2\beta_2}\delta_{\alpha_3\beta_3}
\ea
\item 2 fermionic, 1 bosonic
\ba
\ov(\alpha_1,\beta_1;\alpha_2,\beta_2;a_3,b_3)
&=&O_{000}\; \delta_{\alpha_1\beta_1}\delta_{\alpha_2\beta_2}\delta_{a_3b_3}
+O_{101}\; t^a_{\alpha_1\beta_1}\delta_{\alpha_2\beta_2}T^a_{a_3b_3}
+O_{011}\; \delta_{\alpha_1\beta_1}t^a_{\alpha_2\beta_2}T^a_{a_3b_3}\nn\\
&+&O_{110}\; t^a_{\alpha_1\beta_1}t^a_{\alpha_2\beta_2}\delta_{a_3b_3}
+O_{112}\; t^a_{\alpha1\beta1}t^b_{\alpha2\beta2}
{\cal{P}}_2(a,b;a_3,b_3)\nn\\
&+& O_{111}\; f_{abc}t^a_{\alpha_1\beta_1}t^b_{\alpha_2\beta_2}T^c_{a_3b_3}
\label{2f1b}
\ea
\item 1 fermionic, 2 bosonic
\ba
\ov&=&
O_{000}\; \delta_{\alpha_1\beta_1}\delta_{a_2b_2}\delta_{a_3b_3}+
O_{022}\; \delta_{\alpha_1\beta_1}{\cal{P}}_2(a_2,b_2;a_3,b_3)+
O_{101}\; \delta_{a_2b_2}t^a_{\alpha_1\beta_1}T^a_{a_3b_3}\nn\\
&+&O_{110}\; \delta_{a_3b_3}t^a_{\alpha_1\beta_1}T^a_{a_2b_2}+
O_{111}\; f_{abc}t^a_{\alpha_1\beta_1}T^b_{a_2b_2}T^c_{a_3b_3}
+O_{112}\; t^a_{\alpha_1\beta_1}T^b_{a_2b_2}{\cal{P}}_2(a,b;a_3,b_3)\nn\\
&+&O_{121}\; t^a_{\alpha_1\beta_1}T^b_{a_3b_3}{\cal{P}}_2(a,b;a_2,b_2)
+O_{121}\; t^a_{\alpha_1\beta_1}T^b_{a_3b_3}{\cal{P}}_2(a,b;a_2,b_2)
\nn\\
&+& O_{122}\; f_{abc}t^a_{\alpha_1\beta_1} {\cal{P}}_2(b,d;a_2,b_2){\cal{P}}_2(c,d;a_3,b_3)
\label{1f2b}
\ea
\end{itemize}

The SU(2) symmetry encoded into eqns. (\ref{3f}-\ref{1f2b}) gives various
relations between the {\sl apriori} independent overlap matrix elements,
and therefore between the various cross sections.
For instance in the case
of (\ref{3f}) the overlap has $2^6=64$ values in principle; however there
are only 4 independent projectors, and therefore only 4 independent values.
 The most general relation is the
following: let us give the index assignments:
\be
1=\nu,2=e \;\mbox{  for  fermions         }\qquad
1=W^+,2=W^-,3=W^3\mbox{    for gauge bosons }
\ee
then we have $\sigma_{abc}=\sigma_{a'b'c'}$ where the set $a'b'c'$ is
obtained from $abc$ by the exchange $1\leftrightarrow 2,3\leftrightarrow
3$. So $\sigma_{112}=\sigma_{221},\sigma_{331}=\sigma_{332}$ and so on. It
is easy to realize that the described exchange corresponds to a unitary
transformation with the matrix $i\sigma_2$.
The dressing of the overlap matrix, i.e. resumming leading electroweak
double logs at all orders, is done by applying eq. (\ref{dress}).
The above formula will be useful in order to evaluate  the BN corrections to 
  the high energy cross section for  third family
quarks at  LHC.

\section{Third generation quarks production at LHC}

In this section we will apply some of the  above formulas
for LHC cross sections partially inclusive over soft $W$ and $Z$ emission.
The idea is to analyze  the third family (top and bottom) quark
production at LHC for very large momentum transfer.
In order to give an idea of the size of the corrections we will
also compute  the resummed  Sudakov corrections at leading order. 
In such a
way we can compare  processes without any emission
of $W$ and processes with the same hard final states but with the
possibility to emit soft $W$ bosons.

As is well known we can write the heavy quark production mechanism at LHC
as a convolution of the luminosity functions $L_{ij}$ for the
partons $p_i$ and $p_j$ times the partonic cross sections \cite{web}:
\be\label{sigma}
\frac{d \sigma_{PP\rightarrow Q\bar Q}}{d\hat{s}}=\frac{1}{s}\sum_{i,j}  
L_{ij}(\hat{s})
d\sigma_{p_ip_j\rightarrow Q\bar Q}(\hat{s})\qquad
L_{ij}(\hat{s})=\int_\frac{\hat{s}}{s}^1\frac{dx}{x}
f_{p_i}(x)f_{p_j}(\frac{\hat{s}}{sx})
\ee
where $f_{p_i}(x)$ is the distribution of parton $i$ inside the proton,
$\sqrt{\hat{s}}$ is the partonic c.m. energy and $\sqrt{s}=14$ 
TeV the hadronic one. 
For each channel we will decompose the hard partonic cross sections in 
isospin defined form factors whose EW BN corrections can be directly
computed with eq. (\ref{subs}).

\begin{figure}
      \centering
      \includegraphics[height=50mm] {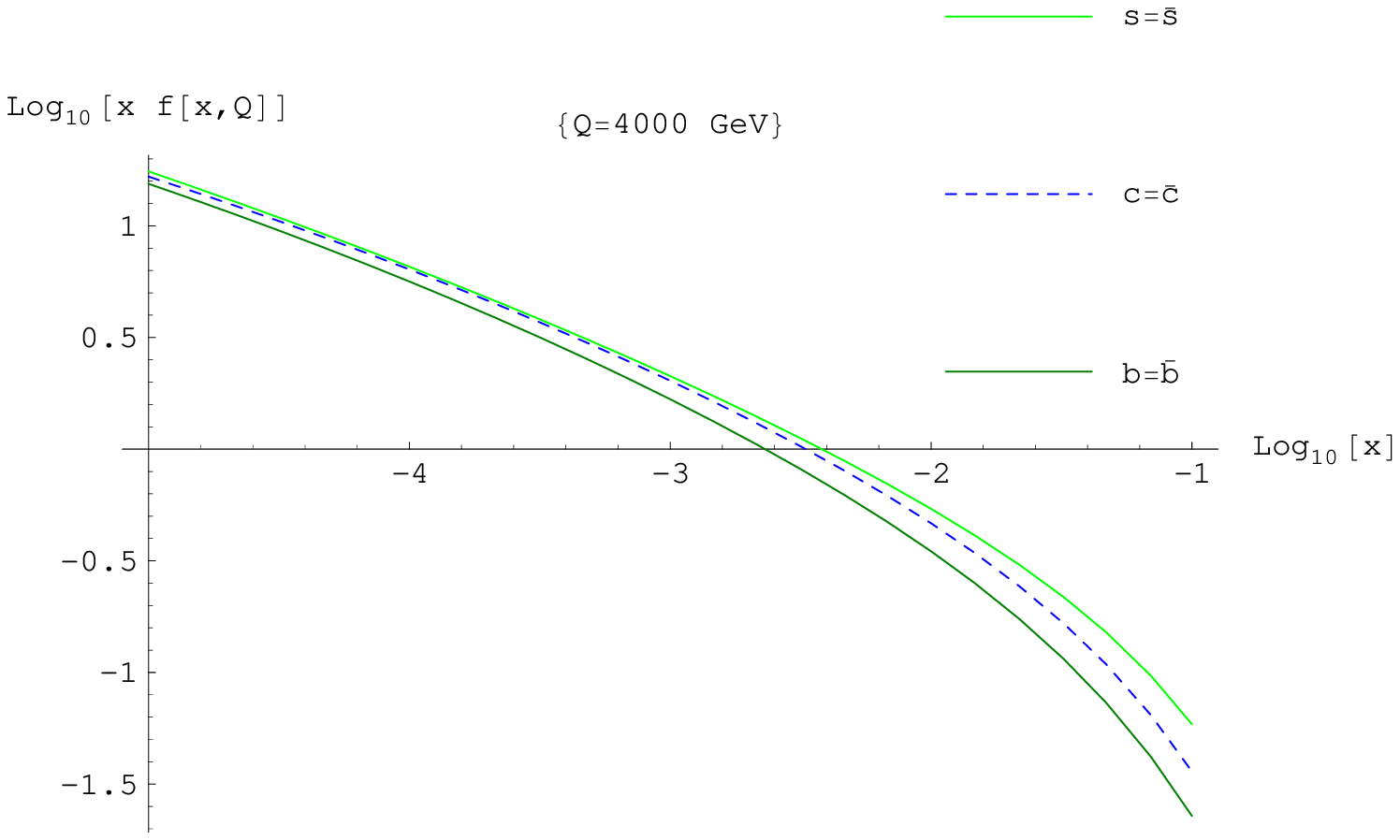}
      \includegraphics[height=45mm] {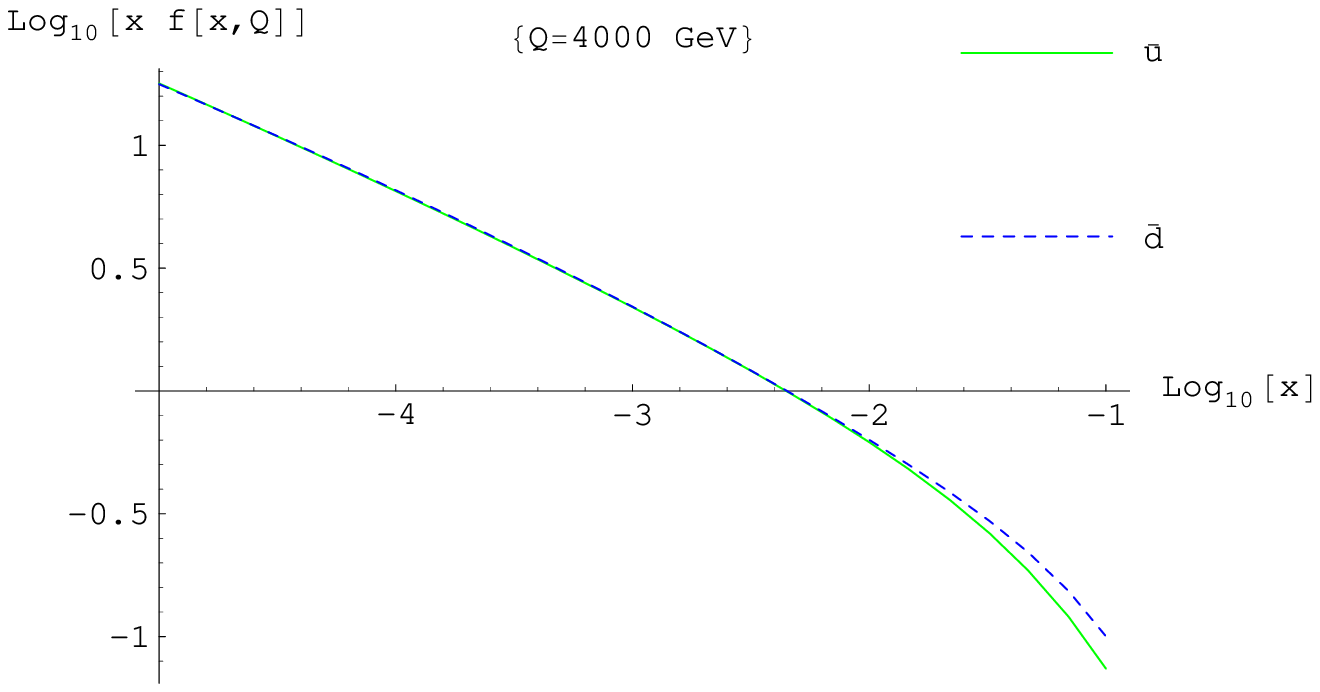}
      \caption{\label{structurefunctions}
Parton distributions for the sea of the proton at $Q=4000 GeV$ as a
function of $x$, the fraction of momentum carried by the parton.
Left: $s=\bar{s},c=\bar{c},b=\bar{b}$. Right: $\bar{u},\bar{d}$. }  
\end{figure}

At this point analyzing the luminosity functions of the sea quarks
of the proton  we can obtain   a quite reasonable  simplification for
the evaluation of the BN corrections to the $q \bar q$ cross section of
eq. (\ref{sigma}):
  the proton sea is approximately an isospin singlet.
In other word the amount of anti up quark inside a proton is almost
the same of the antiquark down and so on for the other
sea families.
This statement, from the SU(2) point of view,
  implies automatically that, with a reasonable 
approximation, the sea quarks of a proton is a flavour singlet state. 
To corroborate our statement we show in fig.\ref{structurefunctions})
the $x$ dependence of 
structure functions of the anti-up and anti-down quarks and of the
remaining sea quarks at fixed energy. This implies that the partonic
process $q\bar{q}\to Q\bar{Q}$ is, with a good approximation, a three leg process and not a
four legs one, since the $\bar{q}$ leg is summed over SU(2) quantum
numbers, and therefore receives no inclusive EW corrections (see also fig. 1).

Let us now turn to the expression for tree level (hard) partonic cross
section  $q\bar{q}\to Q\bar{Q}$, where $ q (Q)$ is a
light (heavy) quark, and we sum over the initial antiquark isospin.
Our notation is 
$\alpha_1=1,2$ for up and  for down type initial quarks,
$\alpha_{2}=1,2$ for $t,\,b$ final states
and
$\alpha_{3}=1,2$ for $\bar t,\,\bar b$ antifermion final states.
The SU(3)$\otimes$SU(2)$\otimes$U(1) couplings are given respectively by
$g_s,g_w,g_y$.
In the massless limit chirality is conserved, so we can label the cross
sections with the chirality of the initial and final fermions: the
chiralities of corresponding antifermions are unambiguously fixed.
The overlap matrix receives various contributions from $s$-channel exchange of 
electroweak gauge bosons and gluons\footnote{We do not consider here t-channel
 contributions from initial sea heavy quarks like $b\bar{t}\to  b\bar{t}$,
a process initiated by the excitations of bottom and top quarks from the
gluon sea.  However for a full  consistent calculation
a careful evaluation of these kind of processes has to be included.}:
\ba \label{qbq1}
\frac{d\sigma^H_{LL}(\alpha_1,\alpha_2,\alpha_3)}{d\cos\theta}&
=&
\frac{\hat{u}^2}{8\pi \hat s^3}
\left[(g^{4}_y\,y_{L\alpha_1}^2\,y_{L\alpha_2}^2+
      2\,g^{2}_y\,g_w^2\,y_{L\alpha_1}\,y_{L\alpha_2}\,t^3_{\alpha_1\alpha_1} \,
t^3_{\alpha_2\alpha_3})\,\delta_{\alpha_2\alpha_3}\right.\\
&&\left.\nonumber
+g_w^4\,t^c_{\alpha_3\alpha_2}(t^ct^d)_{\alpha_1\alpha_1}t^d_{\alpha_2\alpha_3}
+\frac{2}{9} g_s^4 \delta_{\alpha_2\alpha_3} \right]\label{LL}\\
\frac{d\sigma^H_{LR}(\alpha_1,\alpha_2,\alpha_3)}{d\cos\theta}\label{qbq2}
& =&
\frac{\hat{t}^2}{8\pi \hat s^3}
\left[g^{4}_y\,y_{L\alpha_1}^2 \,y_{R{\alpha_2}}^2
+\frac{2}{9} g_s^4  \right] \delta_{\alpha_2\alpha_3}
\\
\frac{d\sigma^H_{RL}(\alpha_1,\alpha_2,\alpha_3)}{d\cos\theta}\label{qbq3}
& =&
\frac{\hat{t}^2}{8\pi \hat s^3}
\left[g^{4}_y\,y_{R\alpha_1}^2 \,y_{L{\alpha_2}}^2  +
\frac{2}{9} g_s^4\right]\delta_{\alpha_2\alpha_3}
\\
\frac{d\sigma^H_{RR}(\alpha_1,\alpha_2,\alpha_3)}{d\cos\theta}\label{qbq4}
& =&
\frac{\hat{u}^2}{8\pi \hat s^3}
\left[g^{4}_y\,y_{R\alpha_1}^2 \,y_{R{\alpha_2}}^2  +
\frac{2}{9} g_s^4 \right]\,\delta_{\alpha_2\alpha_3}
\ea
where $\hat t=-\hat s /2(1-\cos\theta)$ and  $\hat u=-\hat s
/2(1+\cos\theta)$ are the Mandelstam variables in the partonic c.m. frame.
Notice the absence of $g_s^2 g_w^2$  and  $g_s^2 g_y^2$  terms since electroweak and strong amplitudes
do not interfere due to the color structure.
The values of the corresponding hard overlap matrix elements in the LL
channel can be
obtained by equating the general form (\ref{3f}) to the values of the hard
cross sections (\ref{qbq1}, \ref{qbq2}, \ref{qbq3},\ref{qbq4}). Namely, from
$\ov^H_{LL}(\alpha_1,\alpha_1;\alpha_2,\alpha_2;\alpha_3,\alpha_3)
=\frac{d\sigma^H_{LL}(\alpha_1,\alpha_2,\alpha_3)}{d\cos\theta}$ and
defining $\alpha_i=\frac{g_i^2}{4\pi}$ one obtains:
\be\label{prima}
O^H_{000}=\frac{2\,\pi \hat{u}^2}{9 \hat{s}^3}
(\alpha_s^2+\frac{27}{32}\alpha_w^2+\frac{1}{288}\alpha_y^2)
\qquad
O^H_{110}=\frac{\pi \hat{u}^2}{8 \hat{s}^3}(\alpha_w^2+\frac{1}{9}\alpha_w\,\alpha_y)
\ee 
\be
O^H_{011}=\frac{8\,\pi \hat{u}^2}{9 \hat{s}^3}(\alpha_s^2-\frac{9}{32}\alpha_w^2+\frac{1}{288}\alpha_y^2)
\qquad
O^H_{101}=-\frac{\pi \hat{u}^2}{2 \hat{s}^3}(\alpha_w^2-\frac{1}{9}\alpha_w\,\alpha_y)
\ee
The same procedure allows to find the values for the RL channel (in
this case we have two different overlaps: one for the initial  $u_R$
and one for $d_R$, distinguished by the hypercharge contribution):
\be\label{ultima}
O^H_{00}=\frac{2\,\pi \hat{t}^2}{9 \hat{s}^3}
(\alpha_s^2 +\frac{1}{8}\alpha_y^2\,y_{R}^2)
\qquad
O^H_{11}=\frac{8\,\pi \hat{t}^2}{9 \hat{s}^3}(\alpha_s^2 +\frac{1}{8}\alpha_y^2\,y_{R}^2)
\ee
 The dressed overlap matrix
can be now obtained by using the values for the hard overlap matrix of
eqs. (\ref{prima}-\ref{ultima}) and applying the rule of
eq.(\ref{subs}). So for instance 
\be
O_{101}(s)=O^H_{101}\exp[-2L_W]=-\frac{\pi \hat{u}^2}{8 \hat{s}^3}(\alpha_w^2-4 \,\,y_L^2\,\alpha_w\,\alpha_y)
\exp[-2L_W]
\ee
and so on. On the other hand, the $LR,RR$ channels do not receive BN  EW corrections.

The  gluon gluon  hard cross section ($g g\to Q\bar{Q}$)
can be decomposed giving the chirality of the final states:
\be
\frac{ d\sigma^H
}{d \cos \theta}=
\frac{ d\sigma^H_{R}
}{d \cos \theta}+
\frac{ d\sigma^H_{L}
}{d \cos \theta}\quad ;\quad
\frac{ d\sigma^H_{L}
}{d \cos \theta}=\frac{ d\sigma^H_{R}
}{d \cos \theta}
\ee

It is clear that the contributions coming from initial gluons,
from the point of view of $SU(2)$, are 
  two-charged-leg final states  only when left handed heavy quarks
are produced, while the  overlap of the
process $gg\rightarrow Q_{R}\bar Q_R$ is  singlet under   $SU(2)$  decomposition.
The isospin structure of the process $gg\rightarrow Q_{L}\bar Q_L$ is given by 
\be
\frac{ d\sigma_L^H(\alpha_1,\alpha_2)}{d \cos \theta} =
\frac{\pi \alpha_S^2}{4 \hat s}  \left(
\frac{\hat t^2+\hat u^2}{6 \hat t \hat u}-
\frac{3   \left(\hat t^2+\hat u^2\right)}{8 \hat s^2}
\right) \;\delta_{\alpha_1\alpha_2}
\ee
corresponding to an overlap $\ov^H_{}$  given by the coefficients
\ba \label{ovgg}
O^H_{00}=\frac{1}{2}\frac{ d\sigma^H_{L}
}{d \cos \theta}
\quad \quad \quad
O^H_{11}= 2\,\frac{ d\sigma^H_{L}
}{d \cos \theta}  
\ea

The evolved overlap matrix and the respective dressed cross sections
 with the all order resummed virtual and real EW
corrections, can now be obtained by using eqn.(\ref{dress},\ref{subs}) applied to
all the hard overlap form factors. An important feature of such a  channel is the fact that, being a
mixture of s and t-channels, its angular dependence is  different
from the $q \bar q$ s-channel cross section. 
This fact is important not only for $t\bar{t},b\bar{b}$
production but mainly for $t\bar{b},b\bar{t}$ production, where the tree level
$\alpha_W^2$ cross section proceeds only through s-channel annihilation. The outcome is
that the  BN corrected
angular distribution is different from the tree level one (fig. \ref{angtot}).

Finally, in fig. \ref{numerics} we plot the differential cross section for the
process $PP\to b\bar{t}+X$ for the BN and Sudakov cases.

We can obtain a rather simple formula for the BN corrections if we
evaluate  the hard cross section in the limit $g_y,g_w\to 0$; this is a
reasonable approximation since the contributions proportional to $g_s$ 
are the bulk of the
hard cross sections.
It is easy to check  that in this case the same correction is obtained for the
$q\bar q$ and $gg$ channels, allowing  the
factorization of the BN corrections with
respect the  hard QCD cross section:
\ba\label{simple1}
\frac{ d\sigma^{BN}_{}(PP\rightarrow t\bar t)}{d \cos \theta}\approx
\frac{ d\sigma^{BN}_{}(PP\rightarrow b\bar b)}{d \cos \theta} \approx\frac{
  d\sigma^H_{QCD}(PP\rightarrow t\bar t)}{d \cos \theta}\; 
\frac{1}{4}(3+\exp[-2\,L_W(\hat s)])\\\label{simple2}
\frac{ d\sigma^{BN}_{}(PP\rightarrow t\bar b)}{d \cos \theta} \approx
\frac{ d\sigma^{BN}_{}(PP\rightarrow b\bar t)}{d \cos \theta} \approx\frac{
  d\sigma^H_{QCD}(PP\rightarrow t\bar t)}{d \cos \theta}\; 
\frac{1}{4}(1-\exp[-2\,L_W(\hat s)])
\ea

\begin{figure}
      \centering
      \includegraphics[width=15cm] {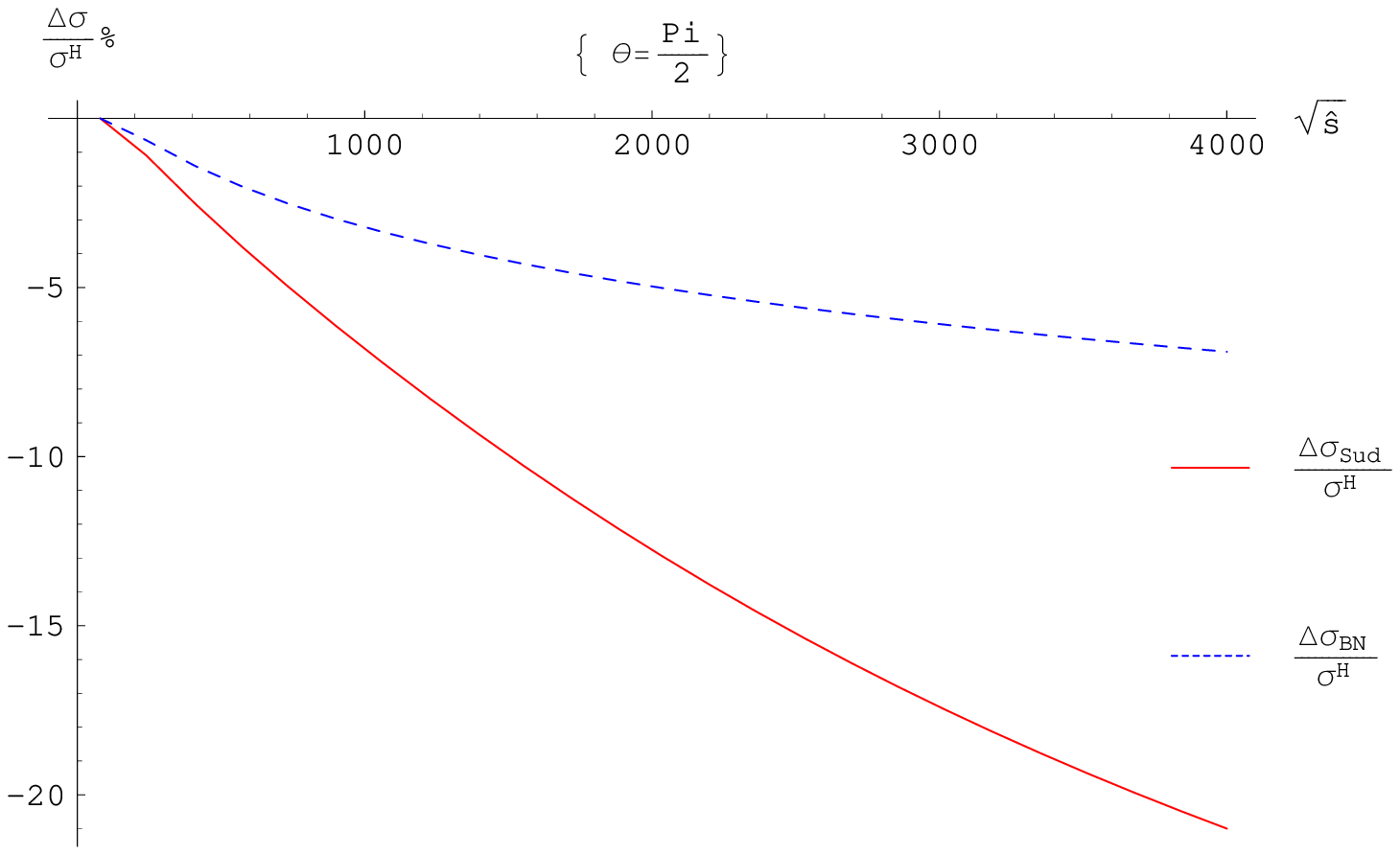}
\includegraphics[width=15cm] {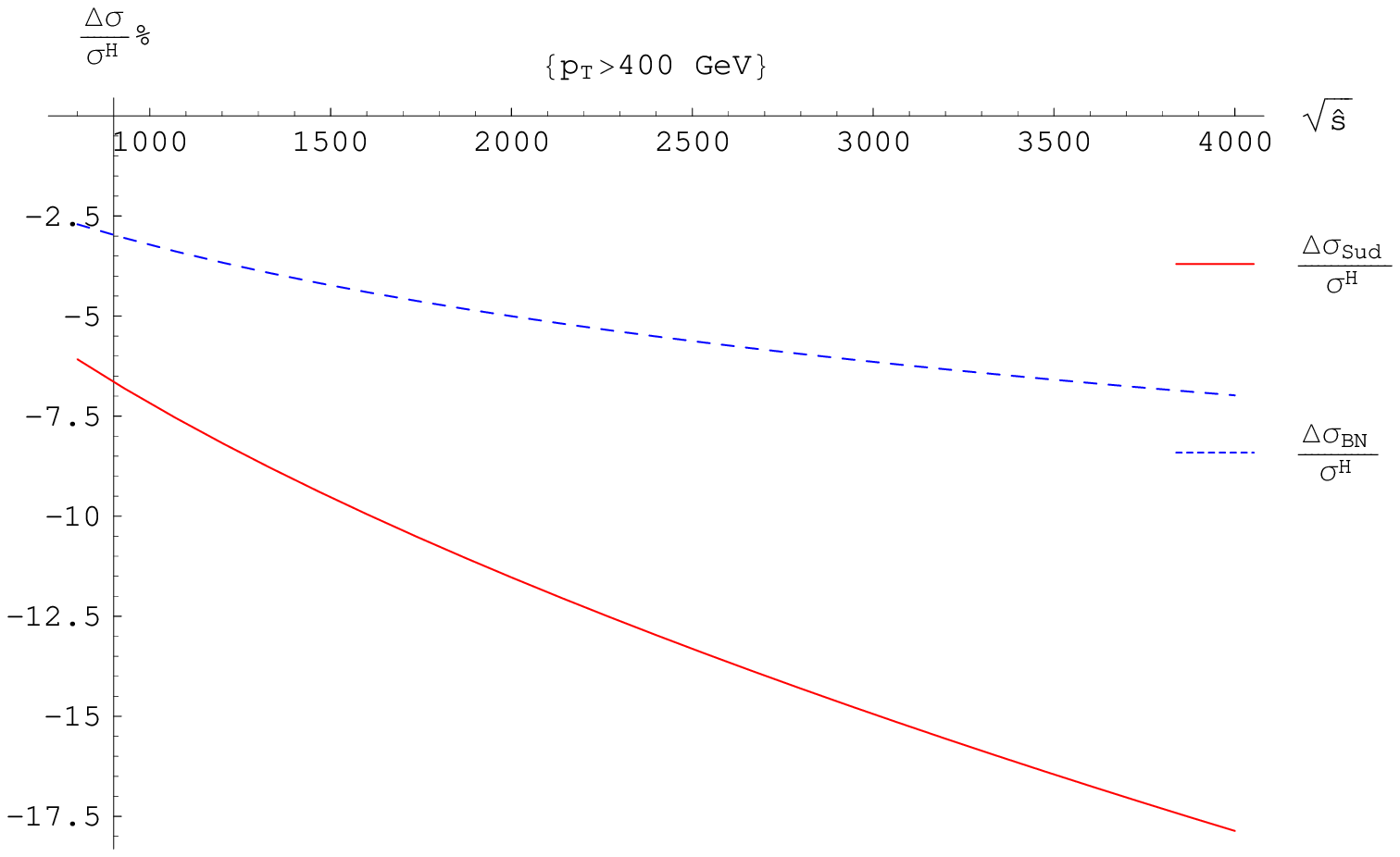}
      \caption{\label{tt12}
Sudakov and BN electroweak corrections at Leading Log order for the
process  $PP\rightarrow t\bar t$ (similar results hold for
  $PP\rightarrow b\bar b$) in the massless limit. Top: effects of radiative corrections on
 $\frac{d^2\sigma}{d\hat{s}d\cos\theta}$ at $\theta=\frac{\pi}{2}$, $\hat{s}$ being the
partonic c.m. energy (which is also the $t \bar{t}$ invariant mass) and
$\theta$ the partonic reference frame scattering angle. Bottom: the same for $
\frac{d\sigma}{d\hat{s}}$, integrated over $\theta$ for $\pt>400$ GeV.
  }  
\end{figure}

\begin{figure}
      \centering
      \includegraphics[width=15cm] {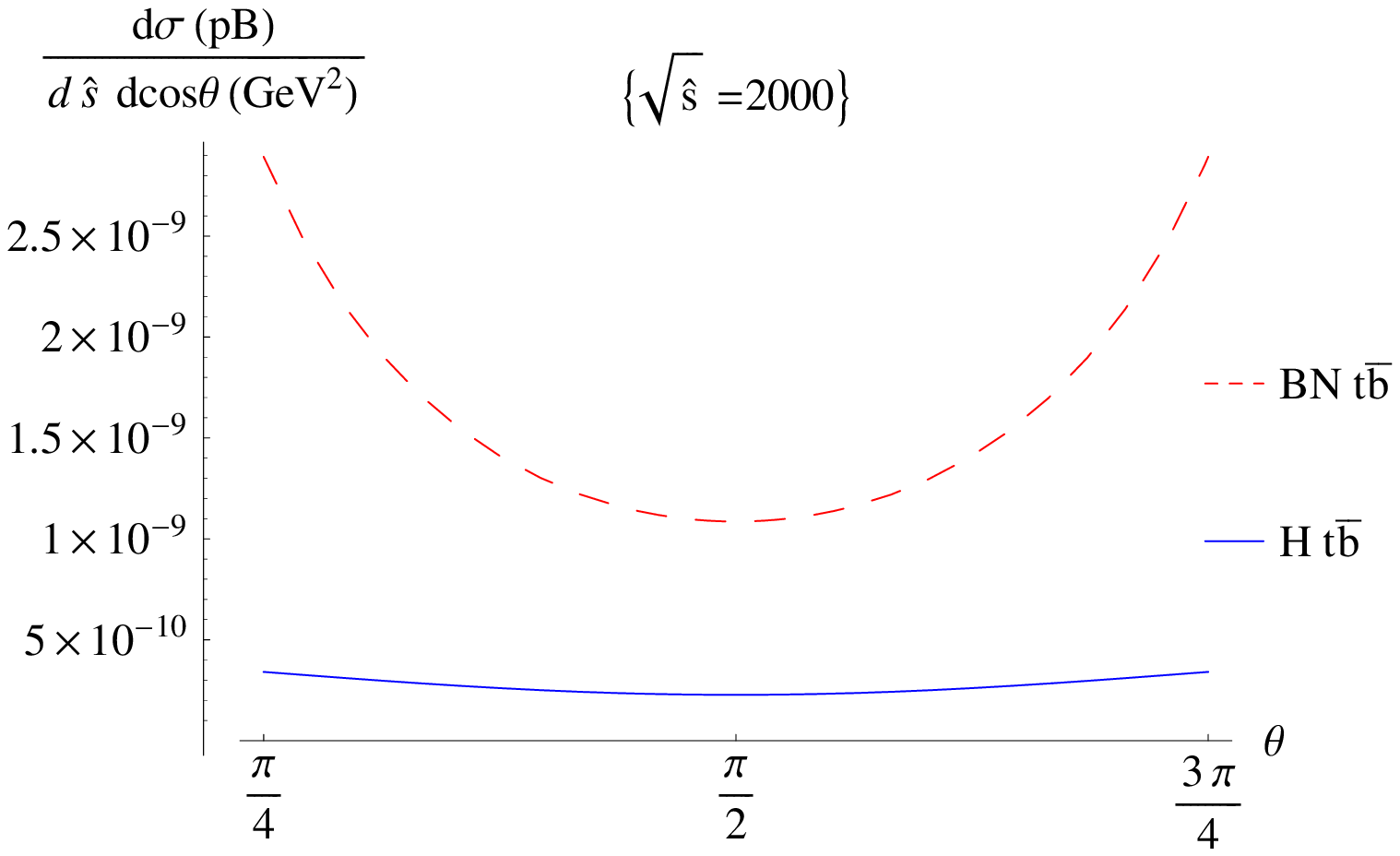}
      \includegraphics[width=15cm] {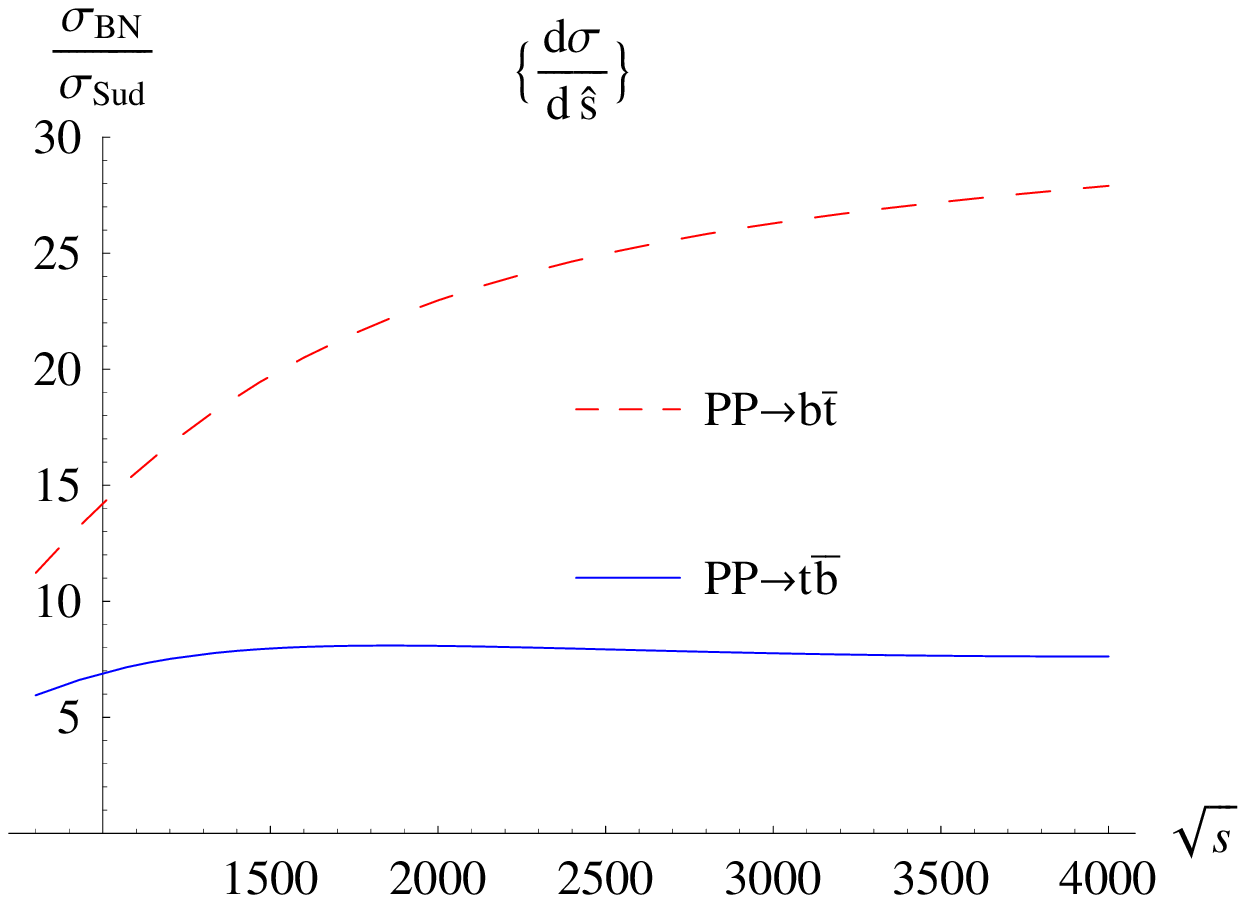} 
          \caption{\label{angtot}
Top: angular dependence at fixed energy ($\sqrt{\hat
  s}=2000$ GeV in the partonic frame) of the LL BN cross
section  (red dashed line) and of the Sudakov one for $t\bar{b}$ production (blue
continuous line).
Bottom: $\sqrt{\hat s}$ dependence of the ratio of BN
  and Sudakov cross sections for $p_{\perp}\geq 400$ GeV. Dashed (red) line for
$b\bar{t}$, continuous (blue) line for $t\bar{b}$.
  }  
\end{figure}

\begin{figure}
      \centering
      \includegraphics[width=15cm] {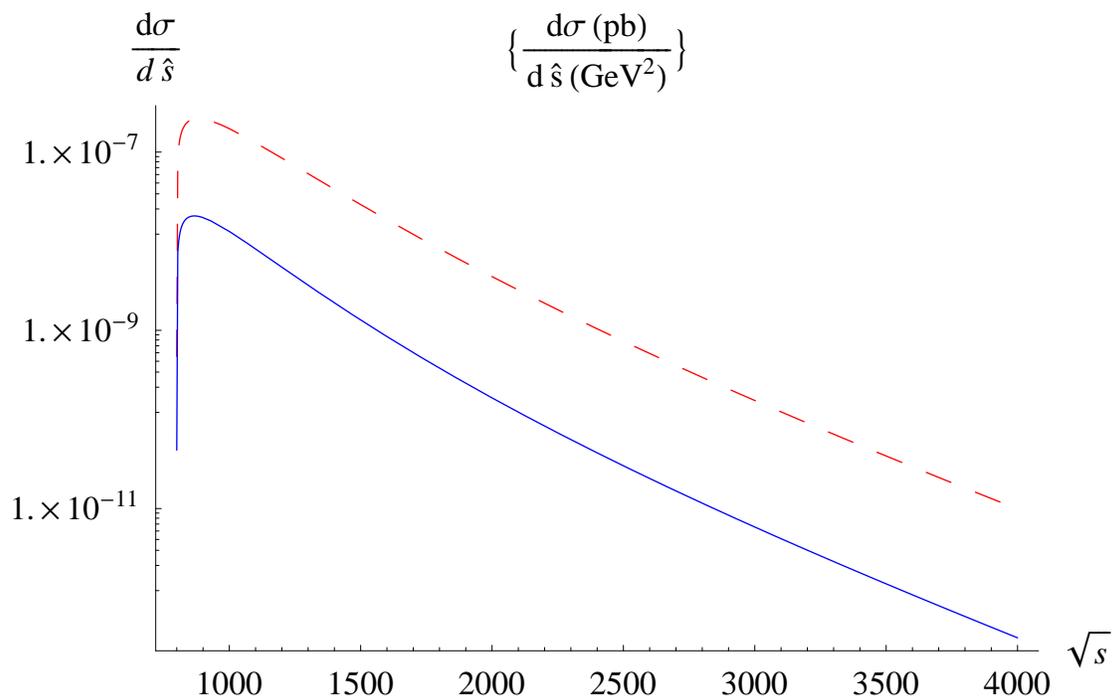}
      \caption{\label{numerics} 
The differential cross section for $b\bar{t}$ production,
$\frac{d\sigma}{d\hat{s}}$ in $\frac{pb}{GeV^2}$, integrated over $\theta$ for $\pt>400$ GeV.
 Dashed (red) line is for the fully inclusive LL BN case, while 
continuous (blue) line for the  exclusive Sudakov case.
  }  
\end{figure}

Let us now comment on our final results for the cross sections
$PP\to Q\bar{Q}+X$ where $Q\bar{Q}=t\bar{t},t\bar{b},b\bar{t},b\bar{b}$, summarized in
figs. \ref{tt12}, \ref{angtot}.
We recall again that  we consider two kinds of
observables:
\ba
``Sudakov''&:(PP\to \mbox{tagged final state}+X )\mbox{ with }
W,Z\notin X \nonumber
\\ \nonumber
``BN''&:(PP\to \mbox{tagged final state} +X) \mbox{ with } W,Z\in X
\ea

Sudakov corrections always depress the tree level cross section \cite{tt}, while in
BN case the sign of the corrections can be positive or negative. 
The results for resummed BN and  Sudakov corrections to
  $t \bar t$ hadronic cross section are
shown in fig. \ref{tt12}. 
Both  corrections are negative in size and more pronounced in the
  Sudakov case.
The $b\bar{b}$ case has a very similar behavior: in fact, in the limit $\alpha_Y\to 0$ 
the  $t\bar{t}$ and $b\bar{b}$ partonic cross sections are equal; the same holds in the
 $t\bar{b}$, $b\bar{t}$ case.
Notice however that the  $t\bar{b}$
and $b\bar{t}$ {\sl hadronic} cross sections are very different, 
due to the different luminosities involved. For instance at
tree level the hadronic cross section for  $t\bar{b}$ depends on
$L_{u\bar{d}}$ while the one for $b\bar{t}$ depends on $L_{d\bar{u}}$,
which is smaller.

In the the $b\bar{t}$,  $t\bar{b}$ channels the BN corrections instead {\sl
  enhance} the cross sections. From
fig. \ref{angtot} we see that the BN enhancement 
 is dramatic: the cross section including
gauge bosons emission is more than one order of magnitude bigger than the exclusive 
(Sudakov) one. This  is due to an interesting interplay
between strong and weak interactions. In fact in these channels, while the
tree level cross sections are proportional to $\alpha_w^2$, when
considering BN corrections they receive big contribution from the strong 
$O(\alpha_s^2)$ channel. 
Moreover,  the tree
 level and the BN cross sections have  different angular behavior
 (see fig. \ref{angtot}).

For 
heavy quark production the leading tree level (with no $W$ emission) cross sections 
are of order  $O(\alpha_S^2)$  when    diagonal   in isospin  
($\sigma^H_{pp\rightarrow t\bar t  },\;\sigma^H_{pp\rightarrow b\bar b  } $),  
while  the two isospin changing one $\sigma^H_{pp\rightarrow b\bar t},
\;\sigma^H_{pp\rightarrow t\bar b}$
 are  $O(\alpha_W^2)$. We can summarize the pattern for the leading tree level and one
 loop EW
 corrected cross sections as follows:
\vspace{0.2cm}

\begin{tabular}{|c||c|c|c|}
\hline
Cross sections & Isospin Structure & Tree Level ($X=0$) & BN corrections\\
\hline
                              & $\delta_{ij}$ & $ \alpha_s^2$ & 
			      $\alpha_s^2\alpha_w\,log^2\hat{s}$\\
$pp\rightarrow Q_i\bar{Q}_j +X$    & & & \\
                              & $i\neq j$  &  $\alpha_w^2 $&  $\alpha_s^2
                         \alpha_w\,log^2\hat{s}$\\
\hline
\end{tabular}

\vspace{0.2cm}
Finally, we expect analogous results for other observables in which nonabelian legs are
detected, such as   single top
production.

\vspace{.2cm}

In this work the CTEQ5M parton distributions \cite{cteq5} have been used.
\section{Conclusions}

The main point of this paper is that emission of real gauge bosons should
be carefully examined when considering  LHC observables. 
As we have seen, including it or
not may result in cross sections differing by an order of magnitude.

As explained in the text, a number of simplifications have been made: we consider the
 proton sea to be an isospin singlet, only resummed double logs are computed and so
 on. Therefore our main results in formulae (\ref{simple1},\ref{simple2}) 
and figs 3,4 have to be taken as first
 order estimates. However first, more detailed calculations are feasible and not too hard
 and second, we think that the outcome is already clear at this preliminary level: by
 considering observables that are inclusive,   
rather than exclusive, of weak bosons emissions, the pattern of radiative electroweak
 corrections changes significantly. In some cases the cross sections that one wants to
 measure are {\sl drammaticaly enhanced} (fig.4), and also differential cross sections
 such as the angular distribution are {\sl significantly different} (fig. 4).

 While
we reckon that it is not entirely clear what can, and will, be measured at the
LHC with respect to ``soft'' final Ws and Zs emission, we think that it is worthwhile 
opening the physics case.
\vspace{0.3cm}

{\bf Acknowledgments}
We thank  M. Ciafaloni for useful theoretical discussions and  M. Grazzini for
help on QCD  Structure Functions.
D.C. acknowledges Cern Theory Division where he started this work.


\def\np#1#2#3{{\sl Nucl.~Phys.\/}~{\bf B#1} {(#2) #3}} \def\spj#1#2#3{{\sl
Sov.~Phys.~JETP\/}~{\bf #1} {(#2) #3}} \def\plb#1#2#3{{\sl Phys.~Lett.\/}~{\bf
B#1} {(#2) #3}} \def\pl#1#2#3{{\sl Phys.~Lett.\/}~{\bf #1} {(#2) #3}}
\def\prd#1#2#3{{\sl Phys.~Rev.\/}~{\bf D#1} {(#2) #3}} \def\pr#1#2#3{{\sl
Phys.~Rep.\/}~{\bf #1} {(#2) #3}} \def\epjc#1#2#3{{\sl Eur.~Phys.~J.\/}~{\bf
C#1} {(#2) #3}} \def\ijmp#1#2#3{{\sl Int.~J.~Mod.~Phys.\/}~{\bf A#1} {(#2) #3}}
\def\ptps#1#2#3{{\sl Prog.~Theor.~Phys.~Suppl.\/}~{\bf #1} {(#2) #3}}
\def\npps#1#2#3{{\sl Nucl.~Phys.~Proc.~Suppl.\/}~{\bf #1} {(#2) #3}}
\def\sjnp#1#2#3{{\sl Sov.~J.~Nucl.~Phys.\/}~{\bf #1} {(#2) #3}}
\def\hepph#1{{\sl hep--ph}/{#1}}

\end{document}